\documentclass[10pt,aps,prd,floatfix,notitlepage,showpacs,nofootinbib,superscriptaddress]{revtex4-1}

\usepackage{amssymb,amsmath}
\usepackage{graphicx}

\begin{document}

\title{The Einstein static universe in Scalar-Fluid theories}

\author{Christian G. B\"ohmer}
\email{c.boehmer@ucl.ac.uk}
\affiliation{Department of Mathematics, University College London, Gower Street, London, WC1E 6BT, UK}

\author{Nicola Tamanini}
\email{nicola.tamanini@cea.fr}
\affiliation{Institut de Physique Th{\'e}orique, CEA-Saclay, F-91191, Gif-sur-Yvette, France}

\author{Matthew Wright}
\email{matthew.wright.13@ucl.ac.uk}
\affiliation{Department of Mathematics, University College London, Gower Street, London, WC1E 6BT, UK}

\date{\today}

\begin{abstract}
A new Lagrangian framework has recently been proposed to describe interactions between relativistic perfect fluids and scalar fields. In this paper we investigate the Einstein static universe in this new class of theories, which have been named Scalar-Fluid theories. The stability of the static solutions to both homogeneous and inhomogeneous perturbations is analysed deriving the relevant cosmological perturbation equations at the linear order. We can find several configurations corresponding to an Einstein static universes which are stable against inhomogeneous perturbations, but unstable against homogeneous perturbations. This shows the possible applications of Scalar-Fluid theories to the inflationary emergent universe scenario.
\end{abstract}

\maketitle

\section{Introduction}

Scalar fields have a prominent role in present cosmology not only since they provide simple inflationary solutions for the early universe, but also for their applications to late-time cosmology. In fact simple scalar field models have been employed to characterize both the inflaton, a hypothetical field introduced to drive the primordial inflationary phase, and dark energy, the entity made responsible for the late-time cosmological acceleration. In general it is believed that further degrees of freedom, beyond the ones of general relativity and Standard Model particles, are needed in order to account for the observations at both early and late times. A scalar field represents thus the simplest way to add just one dynamical degree of freedom into the cosmological framework, and moreover it is usually enough to describe the large-scale effects of high-energy or modified gravity theories, at least at an effective level.

Scalar fields beyond the Standard Model are however expected to possess non negligible interactions with the known matter particles \cite{Carroll:1998zi}, and thus to provide a fifth force deviation from the geodesic motion of freely falling bodies. Because of this fifth force, experiments within the Solar System set stringent constraints on any scalar field model, unless a screening mechanism, such as the well-known chameleon mechanism \cite{Khoury:2003aq}, is introduced. The interaction between the scalar field and the remaining matter sources is commonly characterized by the use of conformal (sometimes disformal) transformations, the case of Scalar-Tensor theories being the most popular. In a recent series of contributions a new framework for coupling a scalar field to matter, including as a sub-class the conformal coupling, has been developed \cite{Boehmer:2015kta,Boehmer:2015sha,Koivisto:2015qua}. This new paradigm uses Brown's Lagrangian formulation of relativistic fluids \cite{Brown:1992kc} to describe the matter sources at a Lagrangian level, providing in such a way new possibilities for coupling the scalar field to the matter sector (see Sec.~\ref{sec:sf_theories}). For this reason, and in analogy with Scalar-Tensor theories, the resulting new class of theories has been dubbed {\it Scalar-Fluid theories} \cite{Koivisto:2015qua}.

The scope of the present paper is to find static cosmological solutions, known as {\it Einstein universes}, within the framework of Scalar-Fluid theories and to analyse their perturbations. This constitutes a simple application of cosmological linear perturbation theory for this newly introduced class of theories, which so far has only been considered for models of interacting dark energy \cite{Boehmer:2015kta,Boehmer:2015sha,Koivisto:2015qua,Tamanini:2015iia}, but in practice it can be applied to other situations, for example primordial inflation. In fact the connection between the Einstein static solution and early universe inflationary theories has been made explicit in the so-called {\it emergent universe scenario}~\cite{Ellis:2002we}, where the initial-time singularity, namely the Big Bang, is replaced by a past asymptotic Einstein universe.

When considering homogeneous and isotropic solutions of the Einstein field equations, one finds that a generic solution will either correspond to an expanding or a contracting universe, the Einstein static universe being the limiting case where the universe does not evolve. However, for this solution to exist in General Relativity, one must introduce the cosmological constant, without it static solutions cannot be found. Following this line of thought, we would expect to find static solutions which are in general unstable with respect to small perturbations which favour a dynamically evolving universe, in agreement with observations. 

The emergent universe paradigm has then stimulated the studies of Einstein static solutions, and their stability under inhomogeneous perturbations, not only in general relativity \cite{Barrow:2003ni}, but also in modified gravity theories representing viable alternatives to single field inflation. In particular the Einstein static universes have been analysed in $f(R)$ theories of gravity \cite{f(R)1, f(R)2, f(R)3}, $f(T)$ gravity \cite{f(T)}, Brans-Dicke theory \cite{Huang:2014fia}, modified Gauss-Bonnet $f(G)$ theories of gravity~\cite{f(G)}, hybrid metric Palatini gravity~\cite{Boehmer:2013oxa}, Einstein-Cartan theory~\cite{Boehmer} and non-constant pressure models \cite{Pressure}. Additionally they have also been investigated in loop quantum cosmology \cite{Loop},  Horava-Lifshitz  gravity~\cite{Horava}, IR modified Horava gravity~\cite{HoravaIR} and non-minimal kinetic coupled gravity \cite{Atazadeh:2015zma}. It is thus interesting to investigate this particular type of solutions within the framework of Scalar-Fluid theories which might constitute alternative models of dark energy and inflation as well.

The paper is organized as follows. In Sec.~\ref{sec:sf_theories} the action of Scalar-Fluid theories will be presented reviewing the main details of the formulation and deriving the relativistic field equations. In Sec.~\ref{sec:algebraic} models with an algebraic coupling between the scalar field and the matter sector will be considered. The cosmological equations will be computed at both background and perturbation levels, while Einstein static solutions will be found for some specific models and their stability will be investigated. In Sec.~\ref{sec:derivative} the same analysis will be applied to Scalar-Fluid models with a derivative coupling between the scalar field and the matter sources. Finally in Sec.~\ref{sec:discussion} the results obtained in the preceding sections will be discussed and the conclusions will be drawn.

Throughout the paper the $(-,+,+,+)$ convention for the signature of the spacetime metric will be used, the speed of light will be set to one $c=1$, and $\kappa^2 = 8\pi G$.

\section{Scalar-Fluid theories: action and field equations}

\label{sec:sf_theories}

In this section we review the variational approach to interacting dark energy that was formulated in~\cite{Boehmer:2015kta,Boehmer:2015sha}. The total action of our interacting dark energy system is 
\begin{align}
  \mathcal{S} = \int \left(\mathcal{L}_{\rm grav} +\mathcal{L}_M+ \mathcal{L}_\phi+ \mathcal{L}_{\rm int}\right) d^4x \,,
\label{010b}
\end{align}
where the gravitational sector $\mathcal{L}_{\rm grav}$ is given by the standard Einstein-Hilbert Lagrangian density
\begin{align}
  \mathcal{L}_{\rm grav} = \frac{\sqrt{-g}}{2\kappa^2}R \,,
\end{align}
with $R$ being the Ricci scalar with respect to the metric $g_{\mu\nu}$, and $g$ denotes its determinant. The Lagrangian density of the scalar field is taken to be of the canonical (quintessence) type
\begin{align}
  \mathcal{L}_\phi = -\sqrt{-g}\, \left[\frac{1}{2}\partial_\mu\phi\,\partial^\mu\phi +V(\phi)\right] \,,
\end{align}
where $V$ is the scalar field potential depending only on $\phi$. The Lagrangian for the relativistic fluid is described using Brown's formulation for $\mathcal{L}_M$  given by \cite{Brown:1992kc,Boehmer:2015kta}
\begin{align}
  \mathcal{L}_M = -\sqrt{-g}\,\rho(n,s) + J^\mu\left(\varphi_{,\mu}+s\theta_{,\mu}+\beta_A\alpha^A_{,\mu}\right) \,,
\label{001b}
\end{align}
where $\rho$ is the energy density of the fluid prescribed as a function of $n$, the particle number density, and $s$, the entropy density per particle. The fields $\varphi$, $\theta$ and $\beta_A$ are all Lagrange multipliers with $A$ taking the values $1,2,3$ and $\alpha_A$ are the Lagrangian coordinates of the fluid. The vector-density particle number flux $J^\mu$ is related to $n$ as
\begin{align}
  J^\mu=\sqrt{-g}\,n\,U^\mu\,, \qquad |J|=\sqrt{-g_{\mu\nu}J^\mu J^\nu}\,, \qquad n=\frac{|J|}{\sqrt{-g}} \,,
\label{056}
\end{align}
where $U^\mu$ is the fluid 4-velocity obeying the relation $U_\mu U^\mu=-1$. 

This just leaves us to determine the interaction Lagrangian $\mathcal{L}_{\rm int}$. We will consider two distinct types of couplings. In~\cite{Boehmer:2015kta} an algebraic coupling between matter and the scalar field was considered. There the interaction Lagrangian took the form
\begin{align}
  \mathcal{L}_{\rm int} = -\sqrt{-g}\, f(n,s,\phi) \,, \label{027}
\end{align}
where $f(n,s,\phi)$ is an arbitrary function which will specify the particular model. In this paper we will consider only one specific type of coupling and take only couplings of the form
\begin{align}
  f(n,s,\phi)=f(\rho,\phi) \,.
\end{align}
This means we will not consider models where the interaction can depend on the entropy density per particle. Moreover, we only consider an implicit dependence on the particle number $n$ through the density $\rho$. Despite these restrictions, this framework is substantial. 

In~\cite{Boehmer:2015sha} a different interaction Lagrangian was considered; a coupling between the matter sector and first derivatives of the scalar field were considered. This time the interacting Lagrangian was given by
\begin{align}
  \mathcal{L}_{\rm int} = f(n,s,\phi) J^\mu\partial_\mu\phi \,,
\label{302b}
\end{align}
where $f$ is again an arbitrary function of the three physical fields. This is the most general coupling term where only one spacetime derivative of the scalar field appears. As with the algebraic coupling, we restrict ourselves to entropy independent interactions and implicit particle number dependence. 

Variation of the total Lagrangian with respect to the metric gives the following Einstein Equations
\begin{align}
  G_{\mu\nu} =\kappa^2\left( T_{\mu\nu} +T_{\mu\nu}^{(\phi)} +T_{\mu\nu}^{\rm (int)} \right)\,,
  \label{EFE}
\end{align}
where the different energy momentum tensors are defined as
\begin{align}
  T_{\mu\nu} & = p\, g_{\mu\nu} + (\rho+p)\, U_\mu U_\nu \,, \\
  T_{\mu\nu}^{(\phi)} &= \partial_\mu\phi\,\partial_\nu\phi -g_{\mu\nu} \left[\frac{1}{2}\partial_\mu\phi\,\partial^\mu\phi +V(\phi)\right] \,, \label{303b} \\
  T_{\mu\nu}^{\rm (int)} & = p_{\rm int}\,g_{\mu\nu} + \left(p_{\rm int}+\rho_{\rm int}\right) U_\mu U_\nu \,.
  \label{030b}
\end{align}
Here the fluid pressure is defined as
\begin{align}
  p = n\frac{\partial\rho}{\partial n}-\rho \,.
\label{eq:111}
\end{align}
In the case of an algebraic coupling, the interacting pressure and energy density are defined as
\begin{align}
  \rho_{\rm int} = f(n,\phi) \quad\mbox{and}\quad p_{\rm int} = n\frac{\partial f(n,\phi)}{\partial n}-f(n,\phi) \,,
\label{024}
\end{align}
whereas in the case of the derivative coupling they are defined as
\begin{align}
  \rho_{\rm int}=0 \qquad\mbox{and}\qquad p_{\rm int} =  -n^2\frac{\partial f}{\partial n} U^\lambda\partial_\lambda\phi \,.
  \label{996}
\end{align}
In what follows we will investigate the Einstein static universe in both of these scenarios. 

\section{Algebraic coupling}
\label{sec:algebraic}

In this section we will consider the Einstein static universe where we assume that the coupling between matter and the scalar field is purely algebraic. This means that the interacting pressure can now be written in terms of $f$ as
\begin{align}
	p_{\rm int} = (\rho +p) \frac{\partial f}{\partial\rho} - f \,.
\end{align}
The cosmological applications for a few particular choices of such an $f$ have been considered in \cite{Boehmer:2015kta}. These models can exhibit a range of interesting cosmological phenomena. Dark energy dominated late time attractors with a dynamical crossing of the phantom barrier have been found, along with scaling solutions, early time matter dominated epochs and a possible inflationary origin. 

\subsection{Background Equations}

To begin with we will consider the background cosmology equations of such models to show that Einstein static universe solutions to the field equations do indeed exist. Let us consider the standard Friedmann-Robertson-Walker (FRW) line element given by
\begin{align}
  ds^2 = -dt^2 + a(t)^2 \left(\frac{dr^2}{1-k r^2} + r^2( d\theta^2+\sin^2\theta \, d\phi^2)\right) \,, \label{FRW}
\end{align}
where $a(t)$ is the cosmological scale factor and $k=-1,0,1$ according to the spatial openness, flatness or closeness of the constant time hypersurfaces, respectively. Inputting this into the Einstein field equations~(\ref{EFE}) we derive the following two Friedmann equations
\begin{align}
  3\frac{k}{a^2}+3H^2 &= \kappa^2\left(\rho +\frac{1}{2}\dot\phi^2 +V +f \right) \,,
  \label{FE1}\\
  \frac{k}{a^2}+2\dot H+3H^2 &=-\kappa^2\left(p+\frac{1}{2}\dot\phi^2 -V +p_{\rm int}\right) \,.
  \label{FE2}
\end{align}
The Klein-Gordon or scalar field equation reduces to
\begin{align}
  \ddot\phi +3H\dot\phi +\frac{\partial V}{\partial\phi} +\frac{\partial\rho_{\rm int}}{\partial\phi} = 0 \,.
\label{KG}
\end{align}

We will now look for an Einstein static universe solution. We set the scale factor $a(t)=a_0={\rm const}$, which implies that $H=\dot{H}=0$. We will also set our scalar field to be a constant: $\phi=\phi_0$, and assume that the perfect fluid obeys a simple linear equation of state $p=w\rho$ where $w$ is a constant lying in the range $-1<w<1$ which is called the equation of state (EoS) parameter. Inputting these assumptions into the two Friedmann equations~(\ref{FE1}) and (\ref{FE2}) yields
\begin{align}
  3\frac{k}{a_0^2} &=\kappa^2\left(\rho_0 +V(\phi_0) +f \right) \label{eqn:Fried1}
  \\
  \frac{k}{a_0^2}&=-\kappa^2\left(p_0-V(\phi_0) +p_{\rm int}\right), \label{eqn:Fried2}
\end{align}
with the Klein-Gordon equation~(\ref{KG}) reducing to
\begin{align}
  V'(\phi_0)+\frac{\partial f}{\partial\phi}\Bigr|_{\phi=\phi_0}=0. \label{KleinGordon}
\end{align}
Combining~(\ref{eqn:Fried1}) and~(\ref{eqn:Fried2}) gives us the simple relation
\begin{align}
\rho_0(1+3w)+f+3p_{\rm int}=2V(\phi_0) \label{constraint}
\end{align}
between the potential and the energy and pressure of both the fluid and the interacting fluid. The above equations give three algebraic equations for the three unknowns $\rho_0$, $a_0$ and $\phi_0$, and thus, as long as $k\neq0$, we can find an Einstein static universe solution. If $k=0$, we note that $a_0$ is undetermined. Note that Eq.~(\ref{KleinGordon}) implies that the static configurations of the scalar field lie in the minima of the effective potential $V(\phi)+f(\rho,\phi)$. Finding an explicit solution will depend on the particular functional form of $f$. We see that, unlike in general relativity, potentially there is also the possibility of a static open universe with $k=-1$ if the function $f$, determining the interacting energy density, is sufficiently negative.

\subsection{Perturbation equations}
We now wish to explore whether the Einstein static universe solutions found above are stable under small perturbations. The perturbed equations for scalar-fluid theories of this type were first derived in~\cite{Koivisto:2015qua}, using a slightly different notation. Here we will review these equations, using our particular functional form for the interacting function $f$. 

We will work with the metric in the Newtonian gauge, also called the longitudinal gauge, which is given by the line element
\begin{align}
  ds^2 = - (1+2\Phi) dt^2+\frac{(1-2\Psi)\, a(t)^2 }{\left[1+\frac{1}{4} k \left(x^2+y^2+z^2\right)\right]^2} \left(dx^2+dy^2+dz^2\right) \,, \label{newtonian}
\end{align}
where both $\Psi$ and $\Phi$ are functions of all the coordinates. Our matter sources can be considered as perfect fluids and so we expect no anisotropic stresses to appear. Hence we may consider the off-diagonal $ij$-components of the Einstein field equations, which indeed read
\begin{align}
  \partial_i\partial_j \left[ \left[1+ \frac{1}{4} k \left(x^2+y^2+z^2\right) \right] (\Psi-\Phi) \right] = 0 \,.
\end{align}
From this equation we immediately find that
\begin{align}
  \Phi = \Psi \,,
\end{align}
as expected since no anisotropies are present in the system. In what follows thus we will simplify all of the equations considering that $\Psi$ equals $\Phi$. This statement is independent of the coupling function $f$ and hence valid for all models in this class. 

We also must determine how the matter variables are perturbed. We perturb the quantities $\rho$, $\phi$, $p$ and $U_\mu$ according to
\begin{align}
  \phi + \delta\phi \,, \qquad \rho + \delta\rho \,, \qquad p + \delta p \,, \qquad U_\mu + \delta U_\mu \,, \label{matterpert}
\end{align}
where $\phi$, $\rho$, $p$ and $U_\mu$ are the background quantities and the perturbation of the four velocity reads
\begin{align}
	\delta U_\mu = \left( -\Psi , \partial_i v \right) \,.
\end{align}
Here $v$ is the scalar perturbation of the matter fluid's velocity. 

We are now in a position to derive the perturbed Einstein equations. We will give the equations directly in Fourier space, so that we write the Laplacian as $\nabla^2 \mapsto -q^2$, where $q$ is the wave number of the fluctuation. For the spatially closed case $k=1$ we have that this wave number must equal $q=n(n+2)$ for positive integer $n=0,1,2,...$, whereas for the spatially open case $k=-1$ we simply have that $q$ is any real number such that $q>1$.

Inserting the perturbed metric into the Einstein equation~(\ref{EFE}), the $00$-component reads
\begin{align}
  \frac{6 k}{a^2} \Psi-\frac{q^2}{a^2} \Psi-4 \pi \left(1+\frac{\partial f}{\partial\rho}\right) \delta\rho -4\pi \left(\frac{\partial f}{\partial\phi}+ V'\right) \delta\phi 
  -8 \pi  \left(\rho + V + f\right) \Psi-3 H \dot\Psi-4 \pi \dot\phi  \dot{\delta\phi} = 0 \,,
  \label{eq:00}
\end{align}
where dots denote differentiation with respect to $t$.
After integrating over $dx^i$, the $0i$-components are
\begin{align}
  8 \pi  (p+\rho) \left(1+\frac{\partial f}{\partial\rho}\right) v  -8 \pi \dot\phi \delta\phi +2 \dot\Psi +2 H \Psi = 0 \,,
\end{align}
which as usual gives the velocity perturbation $v$ in terms of the other perturbed variables. The $ii$-components become, after a simplification using the background equations,
\begin{multline}
	4 \pi  (\rho+p) \frac{\partial^2 f}{\partial\rho^2} \delta \rho +4\pi  \left(1 + \frac{\partial f}{\partial\rho} \right) \delta p +4 \pi  \left[(\rho+p) \frac{\partial^2 f}{\partial\rho\partial\phi} -\frac{\partial f}{\partial\phi} -V'\right] \delta\phi \\
	+4 \pi \dot\phi\dot{\delta\phi} +\left(\frac{k}{a^2}-4\pi\dot\phi^2-2\dot{H}-3H^2\right) \Psi -4 H \dot\Psi -\ddot\Psi = 0 \,.
	\label{eq:ii}
\end{multline}
And finally the perturbation of the scalar field equation~(\ref{KG}) is
\begin{align}
  \ddot{\delta\phi}+3 H \dot{\delta\phi}+\left( \frac{q^2}{a^2} +\frac{\partial^2 f}{\partial\phi^2} +V'' \right) \delta\phi+ \frac{\partial^2 f}{\partial\rho\partial\phi} \delta \rho -2 \left(\ddot\phi+3H\dot\phi\right) \Psi -4 \dot\phi \dot\Psi   = 0 \,.
  \label{eq:scalar}
\end{align}
From equation~(\ref{eq:00}) one can solve for $\delta\rho$ and substitute it into the equations~(\ref{eq:ii}) and~(\ref{eq:scalar}), which will then provide two dynamical equations for the variables $\Psi$ and $\delta\phi$.

We will now insert our background Einstein static universe solution into the perturbed equations above. We will assume adiabatic perturbations, so that the pressure perturbation obeys the same equation of state as the background pressure, $\delta p= w \delta \rho$. Inserting the static solution into~(\ref{eq:00}), we find
\begin{align}
  \frac{3 k}{a_0^2} \Psi-\frac{q^2}{a_0^2} \Psi-4 \pi \left(1+\frac{\partial f}{\partial\rho}\right) \delta\rho= 0 \,.
  \label{eq:01}
\end{align}
The other diagonal equation~(\ref{eq:ii}) simplifies to
\begin{align}
  4 \pi  (w+1)\rho_0 \frac{\partial^2 f}{\partial\rho^2} \delta \rho +4\pi  \left(1 + \frac{\partial f}{\partial\rho} \right) w\delta \rho +4 \pi  \rho_0(w+1) \frac{\partial^2 f}{\partial\rho\partial\phi}  \delta\phi 
  +\frac{k}{a_0^2} \Psi -\ddot\Psi = 0 \,,
  \label{eq:ii3}
\end{align}
and the Klein-Gordon equation~(\ref{eq:scalar}) becomes
\begin{align}
  \ddot{\delta\phi}+\left( \frac{q^2}{a_0^2} +\frac{\partial^2 f}{\partial\phi^2} +V'' \right) \delta\phi+ \frac{\partial^2 f}{\partial\rho\partial\phi} \delta \rho  = 0 \,.
  \label{eq:scalar3}
\end{align}
The off-diagonal equation will not be needed to analyse the stability of the perturbations since the velocity perturbation does not appear in the other equations.

We can solve~(\ref{eq:01}) for $\delta \rho$ and insert this back into~(\ref{eq:ii3}) and~(\ref{eq:scalar3}). We can then write the resulting equations as a coupled two dimensional linear system of second order ordinary differential equations
\begin{align}
  \begin{pmatrix} \ddot{\Psi} \\ \ddot{ \delta \phi} \end{pmatrix} =
  \mathbf{M}
  \begin{pmatrix}  \Psi \\ \delta \phi \end{pmatrix} \,,
  \label{p3}
\end{align}
where $\mathbf{M}$ is a $2\times2$ matrix with constant coefficients involving the background quantities. 

The matrix $\mathbf{M}$ has the following components
\begin{align}
\mathbf{M}=\begin{pmatrix} M_{11} & M_{12} \\ M_{21} & M_{22} \end{pmatrix}
\end{align}
where the components of this matrix are given by
\begin{align}
  M_{11} &=  \left(\frac{(w+1)\rho_0\frac{\partial^2 f}{\partial \rho^2}}{1+\frac{\partial f}{\partial \rho}}+w\right)\left(\frac{3k-q^2}{a_0^2}\right)+\frac{k}{a_0^2} \,, \\
  M_{12} &=  4 \pi  \rho_0(w+1) \frac{\partial^2 f}{\partial\rho\partial\phi} \,, \\
  M_{21} &= - \frac{1}{4\pi(1+\frac{\partial f}{\partial \rho})} \frac{\partial^2 f}{\partial\rho\partial\phi}\left(\frac{3k-q^2}{a_0^2}\right) \,, \\
  M_{22} &=  -\left( \frac{q^2}{a_0^2} +\frac{\partial^2 f}{\partial\phi^2} +V''\right) \,.
  \label{p4}
\end{align}
Since the terms in the matrix are quite involved, there is little hope making generic statement about stability and instability for general $f$.

\subsection{Stability of perturbations}

The linear system of equations described by~(\ref{p3}) is a coupled second order system of differential equations. Therefore it will have four linearly independent solutions as a result of the two eigenvalues of $\mathbf{M}$. Let us denote the eigenvalues of $\mathbf{M}$ by $\lambda_1$ and $\lambda_2$. The solution to the system will involve the frequencies $\pm \sqrt{\lambda_1}$ and $\pm \sqrt{\lambda_2}$, and hence in order for the perturbations to be stable we require the following conditions
\begin{align}
	\Re \lambda_i <0, \quad \Im \lambda_i =0, \quad i=1,2.
\end{align}
Now the components of the matrix $\mathbf{M}$ are too complicated to say anything general about the stability of the Einstein static universe for a generic coupling function $f$. One could attempt using Sylvester's criterion instead of working with the eigenvalues directly, however, the resulting equations are still too involved. Thus we will examine a few specific models corresponding to different functional forms of $f$.

\subsection{Models}

We will consider three separate models assuming different forms for our interaction function $f$. Two of the models were first considered in~\cite{Boehmer:2015kta} where the background cosmological dynamics were analysed. The standard chameleon model~\cite{Khoury:2003aq} can be derived from this interacting Lagrangian approach with algebraic coupling by making the choice for the coupling function $f=-\rho+\rho e^{\beta \kappa \phi}$, and thus we are also able to discuss the stability of the Einstein static universe in this model within this framework. The models we will consider are outlined in the table below
\begin{table}[!ht]
\begin{tabular}{c|c|c}
  \mbox{} & $f$ & $p_{\rm int}$  \\
  \hline
  Model I & $\gamma\,\rho^\alpha\exp(-\beta\kappa\phi)$ & $[\alpha(w+1)-1]f$ \\
  Model II &
  $\gamma\kappa\phi\rho$ & $w f$ \\
  Chameleon field & $-\rho+\rho e^{\beta \kappa \phi}$ & $wf$ 
\end{tabular}
\end{table}

\subsubsection{Model I}
We first consider model I where we take the interacting function $f$ to be a coupling of a power law of the energy density and an exponential in the scalar field, and the potential $V$ to be a standard cosmological exponential potential
\begin{align}
  f(\rho,\phi)=\gamma\,\rho^\alpha e^{-\beta\kappa\phi}, \quad V(\phi)= V_0\, e^{-\lambda\kappa\phi} \,.
\end{align}
The background cosmology of this model was analysed in~\cite{Boehmer:2015kta} for the particular cases of $\alpha=1$ and $\alpha=3$. These models have a range of interesting phenomenology. Late time accelerating attractor solutions were shown to exist for a wide choice of parameters, which can describe dark energy. Scaling solutions were found which may be useful for solving the cosmic coincidence problem, along with solutions undergoing transient inflationary epochs at early times. 

Now let us solve the background equations~(\ref{eqn:Fried1})-(\ref{KleinGordon}) for an Einstein static universe in this model. The Klein Gordon equation gives us the condition
\begin{align}
  \rho_0^\alpha=-\frac{\lambda V_0}{\beta \gamma}e^{(\beta-\lambda)\kappa \phi_0}.
\end{align}
and thus for a positive energy density we require the condition that $\beta \gamma<0$. The Friedmann equations then admit the solution
\begin{align}
  a_0^2=\frac{k \beta(1+3w)}{\kappa^2(1+w)(\beta +\lambda(\alpha-1))V(\phi_0)}
\end{align}
where $\phi_0$ is given implicitly by solving
\begin{align}
  \phi_0=\frac{1}{\beta \kappa}\left(\log\left(\frac{\kappa\beta \gamma }{V'(\phi_0)}\right)-\alpha\log\left(\frac{\kappa\beta(1+3w)}{2\beta \kappa V(\phi_0)+2-3(1+w) \alpha V'(\phi_0)}\right)\right)
\end{align}
Now assuming that both $\lambda, \, \beta\geq0$, requiring that the scale factor is real tells us we will only have a static solution in a closed universe $k=1$ when the EoS paramater lies in the range $w>-1/3$. For the case of an open universe with $k=-1$, the opposite situation arises, with $w$ now lying in the range $-1<w<-1/3$. 

\begin{figure}[!tb]
\includegraphics[trim=0cm 2.5cm 0cm 0cm, clip=true, width=\textwidth]{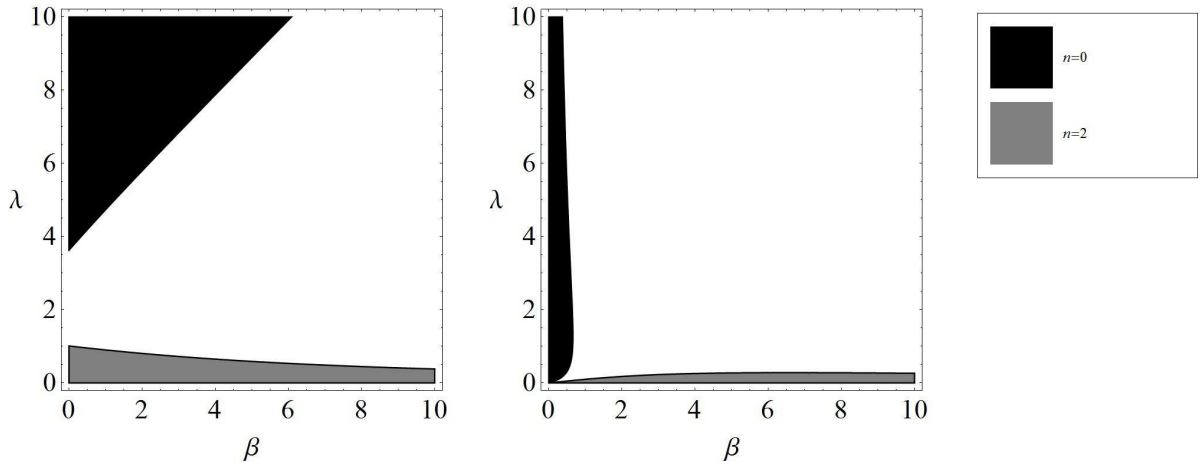}
\caption{Stability of the perturbations in $\beta-\lambda$ parameter space when $w=2/3$. The left panel corresponds to the case $\alpha=1$, the middle panel corresponds to $\alpha=2$. The black region represents the stability of $n=0$ perturbation, the grey region represents the $n=2$ perturbation.}
\label{w13}
\end{figure}

Let us first consider a closed universe, so that we require $w>-1/3$. For the choice of exponents $\alpha=1,2,3,4,5$, it has been checked numerically that the regions of stability of the homogeneous perturbations (corresponding to $q=0$) and the $n=2$ perturbation (corresponding to $q=\sqrt{8}$) do not coincide for any $w$ in the range $-1/3<w<1$.  We can therefore conclude that the Einstein static universe is not stable for these choices of exponent $\alpha$ in this model. We show an example plot of the stability regions of the $n=0$ and $n=2$ perturbations in Fig.~\ref{w13} when the matter EoS $w=1/3$. The grey region indicates the region where the homogeneous perturbations are stable, the grey represents the $n=2$ inhomogeneous perturbation, which overlap nowhere in parameter space. A similar result is found for all values of $w$ in the allowed range. 

Now considering the case of an open universe $k=-1$, we find that for $q$ close to $1$ there are regions of stability. However as we increase $q$ these regions shrink and disappear. This can be seen by noting that for large $q$ it is the case that one of the eigenvalues must be positive. Thus we cannot find a stable static solution in an open universe either.
 
\subsubsection{Model II}

Let us now consider the second model, where we choose the interaction function to be a simple linear coupling between matter and the scalar field
\begin{align}
	f(\rho,\phi)=\gamma \kappa \phi\rho \,.
\end{align}
The background dynamics of this model were also considered in~\cite{Boehmer:2015kta}, assuming a standard exponential potential. The dynamics of this model was very similar to that of the case of general relativity with a canonical scalar field, and thus can in principle replicate the background dynamics of a $\Lambda$CDM universe given a flat enough potential. 

Let us solve the background equations for a static solution. The Klein-Gordon equation~(\ref{KleinGordon}) allows us to solve for the energy density
\begin{align}
  \rho_0=-\frac{1}{\kappa^2\gamma} V'(\phi_0) \,.
\end{align}
Inserting this into the constraint~(\ref{constraint}) yields the following implicit equation for $\phi_0$
\begin{align}
  \kappa\phi_0=-\frac{2\kappa V(\phi_0)}{(1+3w)V'(\phi_0)}-\frac{1}{\gamma} \,.
\end{align}
We can now easily solve one of the remaining Friedmann equations for the scalar factor $a_0$ to give
\begin{align}
  a_0^2=\frac{1}{\kappa^2 V(\phi_0)}\frac{k(1+3w)}{1+w} \,.
\end{align}
And thus positivity of this expression means we only have an Einstein static universe for $k=1$ if $w>-1/3$, and $k=-1$ if $w<-1/3$. To ensure a positive energy density, we require that $V'(\phi_0)<0$ if $\gamma>0$ and $V'(\phi_0)>0$ if $\gamma<0$.  

\begin{figure}[!tb]
\includegraphics[width=0.64\textwidth]{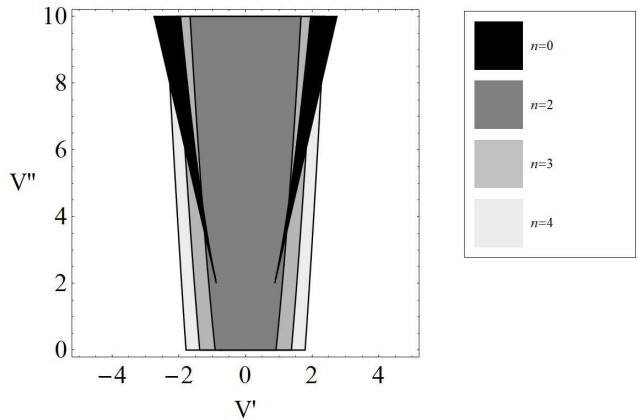}
\caption{Parameter space plot of $V''(\phi_0)$ against $V'(\phi_0)$ showing the regions of stability of the $n=0,2,3,4$ perturbations in Model II with the EoS given by $w=3/4$. There is a small region in which all perturbations are stable. Increasing $n$ increases the region of stability.}
\label{model2stab}
\end{figure}

First we consider the case of a closed universe, $k=1$. Looking at the homogeneous perturbations and the $n=2$, (so that $q=\sqrt{8}$), inhomogeneous perturbations, we find that the regions of stability  of these two overlap for a small range of parameter values. We should note here that one does not need to consider the $n=1$ perturbations, this is simply a gauge degree of freedom. Now as the wave number $n$ increases, the area of stability grows in parameter space, and thus the small region of stability will remain stable to higher wave number perturbations. An example plot showing this behaviour is shown in Fig.~\ref{model2stab} where the $n=0,2,3,4$ perturbations are considered and the EoS was chosen to be $w=3/4$. This region of stability only appears for sufficiently big $w$, for instance the $n=2$ inhomogeneous perturbations are always unstable for $w<1/5$, a result that matches the one obtained with a single sourcing fluid~\cite{Barrow:2003ni}. According to Fig.~\ref{model2stab} the region of stability of inhomogeneous perturbations ($n\geq 2$) is quite large in the $(V,V')$ space. Although only in a small part of such region also the homogeneous perturbations are stable, for the well-known emergent universe scenario only the stability of inhomogeneous perturbations is required in order to have a viable alternative model of inflation \cite{Ellis:2002we,Barrow:2003ni}. Fig.~\ref{model2stab} thus shows that Model~II can be applied consistently to early universe phenomenology within the emergent universe framework.

\begin{figure}[!tb]
\includegraphics[width=0.64\textwidth]{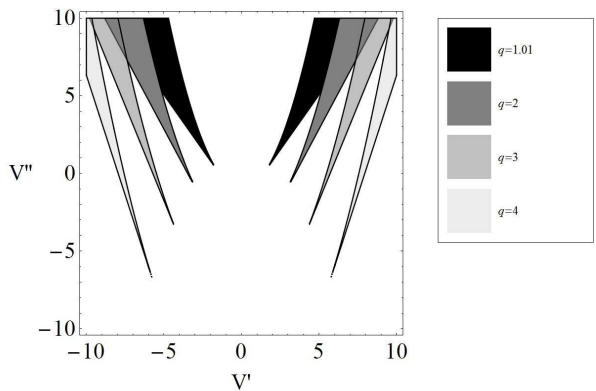}
\caption{Parameter space plot of $V''(\phi_0)$ against $V'(\phi_0)$ showing the regions of stability of the $q=1.01,2,3,4$ perturbations in Model II with the EoS given by $w=-1/2$. The regions do not all overlap, and increasing $q$ eventually makes the stability regions vanish. }
\label{model2stabopen}
\end{figure}

In the case of an open universe $k=-1$, we are unable to find solution which are stable to all perturbations. An example plot is shown in Fig.~\ref{model2stabopen}, where regions of stability for different values of $q$ are shown for the case $w=-1/2$. It is found that increasing $q$ stops the regions of stability overlapping, and for large enough $q$ the stable regions disappear altogether.

\subsubsection{Chameleon Model}

Now let us consider the third of our models. If we choose the interaction function $f$ and the potential $V$ to be
\begin{align}
  f=-\rho+\rho e^{\beta \kappa \phi}, \quad V(\phi)=\frac{M^{4+\alpha}}{\phi^\alpha} \,,
\end{align}
then we recover the standard chameleon mechanism~\cite{Khoury:2003aq} within this scalar-fluid framework. Here $\alpha$ and $\beta$ are positive constants and $M$ is a mass scale.  Such an interaction is of great theoretical interest, since it masks the appearance of the fifth force deviation from general relativity at solar system length scales.

Now looking for static solutions within this model we find the following solution
\begin{align}
  \rho_0&=\frac{2^{\alpha +1} \alpha  M^{\alpha +4} e^{-\frac{1}{2} \alpha(1 +3w)}}{\beta  \kappa } \left(\frac{\beta  \kappa }{\alpha(1 +3w)}\right)^{\alpha+1} \,,
  \\
  a_0^2&= \frac{\beta k }{2^\alpha M^{4+\alpha}\alpha \kappa(1+w)}\left(\frac{\alpha(1 +3w)}{\beta  \kappa }\right)^{\alpha+1}, \quad \phi_0= \frac{\alpha(1 +3w)}{2\beta  \kappa } \,,
\end{align}

Of course, this solution will only exist if $a_0^2$ is positive. Thus depending on the particular parameter choices, there will either exist a $k=+1$ or $k=-1$ static universe. Typically $\alpha$ and $\beta$ are both positive for this model to allow for cosmic acceleration and the screening mechanism,  and thus for a closed $k=1$ static universe we require that $w>-1/3$. There are no open universe solutions unless $w<-1$ so this will not be considered here further.

\begin{figure}[!tb]
\includegraphics[width=0.64\textwidth]{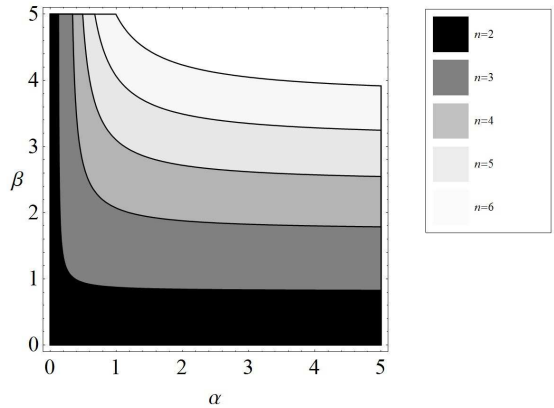}
\caption{Parameter space plot of $\alpha$ against $\beta$ showing the regions of stability of the $n=2,3,4,5,6$ perturbations in the chameleon model with the EoS given by $w=1/3$. The stability regions grow for increasing $n$, and there is a region in which all inhomogeneous perturbations are stable. }
\label{chameleonstable}
\end{figure}

If we analyse the stability matrix of this model, numerically we find that when $\alpha>0$ the homogeneous perturbations are always unstable.  This means one cannot achieve a stable Einstein universe in these chameleon theories in the context of scalar fluid theories. However, analysing the stability of the inhomogeneous perturbations, it is found they are always stable for $\beta$ sufficiently small, see for example Fig.~\ref{chameleonstable}, where the regions of stability of the $n=2,3,4,5,6$ perturbations are plotted for $w=1/3$. With increasing $n$ the stability region grows, and for approximately $\beta<1$ all of the inhomogeneous perturbations are stable. This means that the chameleon model is potentially applicable to the emergent universe framework.

\section{Derivative coupling} 
\label{sec:derivative}

In this section we will analyse the Einstein static universe in the context of a derivative coupling between the matter sector and the scalar field. Such a model was considered in~\cite{Boehmer:2015sha} in the context of dark energy interacting with dark matter.

\subsection{Background equations}

First we will derive the equations governing the background cosmological evolution. As before we will assume the Friedmann-Robertson-Walker metric~(\ref{FRW}). This time the Friedmann equations read
\begin{align}
  3\frac{k}{a^2}+3H^2 &=\kappa^2( \rho +\frac{1}{2}\dot\phi^2 +V) \,,\label{029b}\\
  \frac{k}{a^2}+\left(2\dot H+3H^2\right) &=-\kappa^2( 	p+\frac{1}{2}\dot\phi^2 -V) +n^2\frac{\partial f}{\partial n}\dot\phi \,,
\end{align}
whereas the scalar field equation is modified to
\begin{align}
  \ddot\phi + 3H\dot\phi +V' -n^2\frac{\partial f}{\partial n} 3H =0 \,.
  \label{031b}
\end{align}

Now we look for an Einstein static universe solution in this model, so we assume that our scale factor and all other physical fields are independent of time. The Klein-Gordon equation~(\ref{031b}) contains time derivatives in every quantity except the potential term, so this equation simply reduces to the condition
\begin{align}
  V'(\phi_0)=0 \,,
\end{align}
so that the scalar field of the static universe solution must lie at an extremum of the potential. The static universe is completely independent of the form of the coupling function $f$ at the background level, because the only place at which $f$ enters the field equations it is multiplied by a factor of $\dot{\phi}$. And hence the static solution will be the same as a static solution in standard quintessence. However despite this the equations at the level of the perturbations are different, and so the stability of the static universe should be investigated. The Friedmann equations can easily be seen to reduce to the system
\begin{align}
  \frac{k}{a_0^2} &=\kappa^2 V(\phi_0)\left( \frac{1+w}{1+3w}\right)\,,
  \\ \rho_0 &= \frac{2 V(\phi_0)}{1+3w} \,. \label{eq:bg2} 
\end{align}
In order for the energy density to be positive we will require that the EoS satisfies $w>-1/3$, and this in turn means that for the scale factor to be real we only have a static solution in the case of a closed universe $k=+1$.

\subsection{Perturbations}

We will now derive the general perturbation equations of this derivative coupling model. As before will work in the Newtonian gauge~(\ref{newtonian}) and perturb our matter variables according to~(\ref{matterpert}). Following \cite{Boehmer:2015sha} we will also make an additional assumption on the form of the coupling function $f$, so that the equations are independent of the particle number density $n$ explicitly, with a dependence only implicitly through $\rho$. This leads us to consider the following form of $f$
\begin{align}
  f(n,\phi)= \frac{F(\rho,\phi)}{n} \,.
\end{align}
This means that the interaction energy momentum tensor is given by
\begin{align}
  T_{\mu\nu}^{(\rm int)}=\left( F-(\rho+p)\frac{\partial F}{\partial \rho}\right) U^{\lambda}\partial_{\lambda} \phi (g_{\mu\nu}+U_{\mu \nu}) \,.
\end{align}

Once again looking at the off-diagonal $ij$-components of the field equations immediately gives
\begin{align}
  \Phi = \Psi \,,
\end{align}
since once again no anisotropies are present. Thus as is the algebraic coupling case, in what follows we will simplify the equations considering that $\Phi$ equals $\Psi$ and we will give the equations directly in the Fourier space: $\nabla^2 \mapsto -q^2$.

The $00$-component of the Einstein field equations reads
\begin{align}
  \left(\frac{6 k}{a^2} -\frac{q^2}{a^2}-\kappa^2\left(\rho + V \right) \right) \Psi-3 H \dot\Psi-\frac{\kappa^2}{2} ( \delta\rho + V' \delta\phi 
  + \dot\phi  \dot{\delta\phi}) = 0 \,,
  \label{eq:10}
\end{align}
which is independent of the coupling function $f$, exactly as in the background case. The $0i$-component are (after integrating over $dx^i$)
\begin{align}
  \kappa^2 \left( F \dot{\phi}+ (p+\rho) \left(1-\dot{\phi}\frac{\partial F}{\partial\rho} \right) \right) v  -\kappa^2 \dot\phi \delta\phi +2 \dot\Psi +2 H \Psi = 0 \,.\label{eq:11}
\end{align}
The $ii$-components are, after a simplification using the background equations
\begin{multline}
  \ddot\Psi+4 H \dot\Psi +\left(2\dot{H}+3H^2-\frac{k}{a^2}+\frac{\kappa^2}{2}(\dot\phi^2+F\dot{\phi}-(\rho+p)\frac{\partial F}{\partial\rho}\dot{\phi})\right) \Psi \\
  +\frac{\kappa^2}{2}  (\rho+p) \dot{\phi} \frac{\partial^2 F}{\partial\rho^2} \delta \rho + \frac{\kappa^2}{2}  \left( \dot{\phi}\frac{\partial F}{\partial\rho}-1 \right) \delta p \\+ \frac{\kappa^2}{2} \left[(\rho+p)\dot{\phi} \frac{\partial^2 F}{\partial\rho\partial\phi}-\dot{\phi}\frac{\partial F}{\partial\phi} +V'\right] \delta\phi 
  + \frac{\kappa^2}{2} ((\rho+p)\frac{\partial F}{\partial\rho}-F-\dot\phi)\dot{\delta\phi}   = 0 \,.
  \label{eq:ii2}
\end{multline}
And finally the perturbation of the scalar field equation reads
\begin{multline}
  \left[3(F-(\rho+p)\frac{\partial F}{\partial\rho})+4 \dot\phi \right]\dot{\Psi}
  + \left(2\ddot\phi+6H\dot\phi-3H( (\rho+p)\frac{\partial F}{\partial\rho}-F) \right) \Psi \\
  +3H (\rho+p) \frac{\partial^2 F}{\partial\rho^2} \delta \rho+3H\frac{\partial F}{\partial\rho} \delta p  - ((\rho+p)\frac{\partial F}{\partial\rho}-F)\frac{q^2}{a^2} v \\
  +\left( -\frac{q^2}{a^2} +3H((\rho+p)\frac{\partial^2 F}{\partial\rho\partial \phi}-\frac{\partial F}{\partial\phi})  -V'' \right) \delta\phi	-3 H \dot{\delta\phi}-\ddot{\delta\phi} = 0 \,.
  \label{eq:scalar2}
\end{multline}

\subsection{Stability of the static universe}

Now let us insert our Einstein static universe solution into the perturbation equations. The $00$ equation~(\ref{eq:10}) now becomes
\begin{align}
  \left( \frac{3k}{a_0^2}-\frac{q^2}{a_0^2}\right)
  \Psi=\frac{\kappa^2}{2}\delta\rho \,. \label{densitypert}
\end{align}
The $0i$-component~(\ref{eq:11}) takes the particularly simple form
\begin{align}
  \kappa^2 \rho_0(1+w) v=-2\dot{\Psi}\,, \label{velocpert}
\end{align}
which allows one to find the velocity perturbation easily in terms of the metric perturbation. The $ii$ components~(\ref{eq:ii2}) reduce to
\begin{align}
  \frac{\kappa^2}{2}(-\delta p +((\rho+p)\frac{\partial F}{\partial\rho}-F )\dot{\delta \phi})-\frac{k}{a_0^2}\Psi+\ddot{\Psi}=0\,, \label{eq:13}
\end{align}
while the scalar field equation~(\ref{eq:scalar2}) becomes
\begin{align}
  \ddot{\delta \phi} +\left( \frac{q}{a_0^2}  + V'' \right)\delta \phi+3((\rho+p)\frac{\partial F}{\partial\rho}-F)\dot{\Psi}-(F-(\rho+p)\frac{\partial F}{\partial\rho})\frac{q^2}{a_0^2} v=0 \,. \label{eq:14}
\end{align}

Once again we will now assume an adiabatic perturbation, so that $\delta p= w \delta \rho$.  Substituting the density perturbation from~(\ref{densitypert}) into~(\ref{eq:13}) and the velocity perturbation~(\ref{velocpert}) into~(\ref{eq:14}) the system of equations reduce to the following two dimensional system 
\begin{align}
  \frac{\kappa^2}{2}\left(\rho_0(w+1)\frac{\partial F}{\partial\rho}-F\right)\dot{\delta \phi}+\left(\frac{w q^2}{a_0^2}-\frac{(3w+1)k}{a_0^2}\right) \Psi+\ddot{\Psi}&=0 \,,
  \label{stat1} \\
  \ddot{\delta \phi} +\left( \frac{q}{a_0^2}  + V'' \right)\delta \phi +\left(3+\frac{2q^2}{\kappa^2 a_0^2 (1+w)\rho_0}\right)\left(\rho_0(w+1)\frac{\partial F}{\partial\rho}-F\right)\dot{\Psi}&=0 \,.
  \label{stat2}
\end{align}

Now let us introduce the following vector
\begin{align}
  \boldsymbol{X}= \begin{pmatrix}  \Psi \\ \delta \phi \end{pmatrix} \,,
\end{align}
which means we can write the above system of equations~(\ref{stat1}), (\ref{stat2}) as the following two dimensional matrix equation
\begin{align}
  \boldsymbol{\ddot{X}}+A \boldsymbol{\dot{X}}+B \boldsymbol{X} =0 \,. \label{matrixeqn1}
\end{align}
Here the matrices $A$ and $B$ have been defined as
\begin{align}
  A=  \begin{pmatrix} 0  & \frac{\kappa^2}{2}P \\ (3+\frac{2q^2}{\kappa^2a_0^2 (1+w)\rho_0})P & 0 \end{pmatrix}\,, \quad 
   B=  \begin{pmatrix} \frac{w q^2}{a_0^2}-\frac{(3w+1)k}{a_0^2}  & 0 \\ 0 & \left( \frac{q^2}{a_0^2}  + V'' \right) \end{pmatrix}\,,
\end{align}
where we have introduced the quantity $P$
\begin{align}
  P=\rho_0(w+1)\frac{\partial F}{\partial\rho}-F \,.
\end{align}

Now to reduce the equation to a first order system, we introduce the vector
\begin{align}
  \boldsymbol{Y}=\boldsymbol{\dot{X}} \,,
\end{align}
so that the equation~(\ref{matrixeqn1}) can be written as the following first order autonomous system
\begin{align}
  \begin{pmatrix}  
    \boldsymbol{\dot{Y}} \\ \boldsymbol{\dot{X}} 
  \end{pmatrix}= 
  \begin{pmatrix}
    -A & -B \\ I_2 & 0
  \end{pmatrix} 
  \begin{pmatrix}
    \boldsymbol{Y} \\ \boldsymbol{X}
  \end{pmatrix}\,, \label{firstordersyst}
\end{align}
where $I_2$ denotes the $2\times 2$ identity matrix. For the system to be stable we simply require that the eigenvalues of the above system are purely imaginary. Let us write
\begin{align}
  A=  \begin{pmatrix} 0  & a_1\\ a_2 & 0 \end{pmatrix}
  \,, \quad 
  B=  \begin{pmatrix} b_1  & 0 \\ 0 & b_2 \end{pmatrix} \,.
\end{align}
Then the four eigenvalues of the system~(\ref{firstordersyst}) in terms of $a_i$ and $b_i$ are simply
\begin{align}
  \lambda_i= \pm \frac{1}{\sqrt{2}} \sqrt{a_1 a_2-b_1-b_2 \pm \sqrt{(b_1+b_2-a_1a_2)^2-4 b_1 b_2}} \,. \label{eigen}
\end{align}

Now we immediately see from the definitions of $a_1$ and $a_2$ that $a_1 a_2\geq 0$. Thus requiring the eigenvalues~(\ref{eigen}) are imaginary reduces to the following conditions
\begin{align}
  b_1>0 \,, \quad b_2>0 \,,
\end{align}
\begin{align}
  a_1>0, \, a_2>0, \, a_1 a_2<  b_1+b_2-2\sqrt{b_1 b_2}, \quad {\rm or}\quad  a_1<0, \, a_2<0, \, a_1 a_2<  b_1+b_2+2\sqrt{b_1 b_2}.
\end{align}

For the homogeneous perturbations we set $q=0$. Then requiring $b_1>0$ tells us we must have $w<-1/3$, and $b_2>0$ tells us the potential must lie at a minimum $V''(\phi_0)>0$.  However, we have already seen that the static universe solution requires that $w>-1/3$, and hence the homogeneous perturbation cannot be stable. And thus no static universe will be homogeneously stable in these derivative coupled models. 

Now let us examine the stability against inhomogeneous perturbations. $b_1>0$ will be satisfied as long as $q^2>4$ and $w>1/(q^2-3)$ (so $w>1/5$ for the $n=2$ perturbation \cite{Barrow:2003ni}). The first of these condition is always satisfied since the smallest inhomogeneous perturbation is the $n=2$ mode, which corresponds to $q^2=8$.  The condition $b_2>0$ will be satisfied as long a we are at a minimum of the potential: $V''(\phi_0)>0$.  We can simplify the quantity $a_2=(3+q^2)P$. Therefore the condition for stability is:
\begin{align}
  \frac{\kappa^2}{2}(3+q^2)P^2< \left( \sqrt{\frac{q^2}{a_0^2}  + V''} +\frac{\sqrt{w q^2-(3w+1)}}{a_0}\right)^2 \qquad \text{if } P>0 \,, \label{positiveP}
\end{align}
and
\begin{align}
  \frac{\kappa^2}{2}(3+q^2)P^2< \left(\sqrt{\frac{q^2}{a_0^2}  + V''} - \frac{\sqrt{w q^2-(3w+1)}}{a_0}\right)^2 \qquad \text{if } P<0 \,. \label{negativeP}
\end{align}
These conditions can be satisfied for all inhomogeneous perturbations. To show this we will derives some sufficient conditions for these inequalities to be true. For positive $P$ a sufficient condition for~(\ref{positiveP}) to be true is
\begin{align}
  \frac{\kappa^2}{2}(3+q^2)P^2<\frac{q^2}{a_0^2} \,,
\end{align}
so we need
\begin{align}
  \frac{\kappa^2}{2} a_0^2 P^2 < \frac{q^2}{3+q^2}, \quad {\textrm {for all }} q>\sqrt{8} \,,
\end{align}
which is satisfied if
\begin{align}
  \frac{\kappa^2}{2} a_0^2 P^2 < \frac{8}{11} \,,
\end{align}
which alternatively we can write as
\begin{align}
  \left( \frac{1+3w}{1+w}\right)\frac{P^2}{2V(\phi_0)} < \frac{8}{11} \,.
 \label{eq:cond_1}
\end{align}
For $P<0$ a sufficient condition for~(\ref{negativeP}) to be true is for
\begin{align}
  \frac{\kappa^2}{2}(3+q^2)P^2<\frac{(1-\sqrt{w})^2q^2}{a_0^2} \,,
\end{align}
holding when $0<w<1$. This can then be reduced to the following condition which will ensure the stability for all inhomogeneous perturbations
\begin{align}
  \left( \frac{1+3w}{(1+w)(1-\sqrt{w})^2}\right)\frac{P^2}{2V(\phi_0)} < \frac{8}{11} \,.
 \label{eq:cond_2}
\end{align}
This is a stricter condition than for positive $P$. Nonetheless it can still easily be satisfied for sufficiently small $P$. 

Let us examine the form of $P$ for different choices of coupling functions. If $F$ of the form $F={\rm const}$ or $F=F(\phi)$, then we simply have $P=-F$. If $F$ is a constant then the stability of inhomogeneous perturbations will depend not only on the value of $F$, but also on $V(\phi_0)$ according to the conditions (\ref{eq:cond_1}) and (\ref{eq:cond_2}). An interesting case is $F(\phi) = \xi \sqrt{V(\phi)}$, whose background cosmology curiously results to be equivalent to the one analysed in \cite{Tamanini:2014mpa} in the context of $k$-essence, as shown in \cite{Boehmer:2015sha}. According to the conditions (\ref{eq:cond_1}) and (\ref{eq:cond_2}), in this situation we find that the stability of inhomogeneous perturbations will no longer depend on the scalar field potential, but only on the constant $\xi$. For example considering $w=1/3$, which is expected for early universe applications, we will find stability approximately if $-0.416 < \xi < 0.985$. A similar reduction applies in the case $F=\gamma \sqrt{\rho}$, which has been studied in \cite{Boehmer:2015sha}. In this case using the background equation (\ref{eq:bg2}) the conditions (\ref{eq:cond_1}) and (\ref{eq:cond_2}) become again independent of the scalar field and the stability of perturbations will be determined by the constant $\gamma$. For early universe applications ($w=1/3$) one finds stability of the inhomogeneous perturbations if approximately $-1.25 < \gamma < 9.80$.

These last examples shows that Scalar-Fluid models with derivative couplings can easily be used in the context of the emergent universe scenario where the stability of inhomogeneous perturbations and the instability of the homogeneous perturbations are required for the viability of this alternative model of inflation.

\section{Discussion}
\label{sec:discussion}

In this work we have analysed Einstein static universe solutions in the newly proposed framework of Scalar-Fluid models, where an interaction between an effective perfect fluid and a scalar field is introduced directly at the level of the action. We have shown that generically static solutions exist and we have studied their stability against both homogeneous and inhomogeneous perturbations, deriving the relevant cosmological perturbation equations at the linear level.

In the case of purely algebraic couplings, we are unable to find simple analytic conditions to determine the stability of our static solutions. The reason for this is the complicated structure of the matrix whose eigenvalues determine the stability properties of the perturbed solutions. We were thus forced to consider individual models and numerically explore the regions of stability. We have analysed three particular models. The first of these, assuming a non-linear exponential coupling, was shown to be generically unstable, to both homogeneous and inhomogeneous perturbations. However when a simple linear coupling is considered, namely $\rho_{\rm int} \propto \phi \rho$, it is found that there is a small region of parameter space where the Einstein static universe is stable to both homogeneous and inhomogeneous perturbations, while there is a large region in parameter space where it is stable only against inhomogeneous perturbations. This last situation is exactly the one required by the emergent universe paradigm, implying that such model can be applied to early universe phenomenology as an alternative inflationary scenario. Similar results can be found with the third Scalar-Fluid coupling, which reproduces the well-known chameleon mechanism. This incidentally suggests possible applications of screening models to the emergent universe scenario which could be taken into account for future analyses.

On the other hand, when considering an arbitrary coupling between the fluid's four velocity and the derivative of the scalar field, we are able to make some generic statements applicable to all models of this type. It is found that homogeneous perturbations are always unstable, while the stability of inhomogeneous perturbations is determined by simple inequalities (see Eqs.~(\ref{eq:cond_1}) and (\ref{eq:cond_2})) depending on the scalar field potential and the specific form of the Scalar-Fluid derivative coupling. Particular models of this kind, the ones admitting stable inhomogeneous perturbations, are thus suitable for applications within the context of the emergent universe scenario.

In general thus the results obtained in this work show that Scalar-Fluid theories might well constitute new interesting inflationary paradigms. Their broad applications to early universe phenomenology, also as possible mechanisms of reheating, deserves to be studied in future works. It thus appears that these theories offer the possibility of studying early time and late time phenomenology using a single model based on a well defined Lagrangian approach.

\end{document}